\documentclass[journal=jacsat,manuscript=article]{achemso}

\author{Arnold O. Benz}
\email{benz@astro.phys.ethz.ch}
\affiliation[ETH Zurich]
{Institute for Astronomy, ETH Zurich, 8093 Z\"urich, Switzerland}
\author{Simon Bruderer}
\affiliation[MPE Garching]{Max Planck Institut f\"{u}r extraterrestrische Physik, Giessenbachstrasse 1, 85748 Garching, Germany}
\author{Ewine F. van Dishoeck}
\affiliation[Leiden Observatory, MPE Garching] {Leiden University, PO Box 9513, 2300 RA Leiden, The Netherlands}
\author{Pascal St\"auber}
\affiliation[ETH Zurich]{Institute of Astronomy, ETH Zurich, 8093 Z\"urich, Switzerland}
\author{Susanne F. Wampfler}
\affiliation[ETH Zurich, CSPF Copenhagen]{Centre for Star and Planet Formation, Natural History Museum of Denmark,University of Copenhagen, \O{}ster Voldgade 5-7, DK-1350 K\o{}benhavn K, Denmark}

\title[\texttt{achemso} demonstration]
{Neutral and Ionized Hydrides in Star-forming Regions --
Observations with Herschel/HIFI\footnote{{\it Herschel} is an ESA space observatory with science instruments provided by European-led Principal Investigator consortia and with important participation from NASA.} }
\begin{document}
%%%%%%%%%%%%%%%%%%%%%%%%%%%%%%%%%%%%%%%%%%%%%%%%%%%%%%%%%%%%%%%%%%%%%
%% The manuscript does not need to include \maketitle, which is
%% executed automatically.  The document should begin with an
%% abstract, if appropriate.  If one is given and should not be, the
%% contents will be gobbled.
%%%%%%%%%%%%%%%%%%%%%%%%%%%%%%%%%%%%%%%%%%%%%%%%%%%%%%%%%%%%%%%%%%%%%
\begin{abstract}
  The cosmic abundance of hydrides depends critically on high-energy UV, X-ray, and particle irradiation. Here we study hydrides in star-forming regions where irradiation by the young stellar object can be substantial, and density and temperature can be much enhanced over interstellar values. Lines of OH, CH, NH, SH and their ions OH$^+$, CH$^+$, NH$^+$, SH$^+$, H$_2$O$^+$, and H$_3$O$^+$  were observed in star-forming regions by the HIFI spectrometer onboard the {\it Herschel Space Observatory}. Molecular column densities are derived from observed ground-state lines, models, or rotational diagrams. We report here on two prototypical high-mass regions, AFGL 2591 and W3 IRS5, and compare them to chemical calculations making assumptions on the high-energy irradiation. A model assuming no ionizing protostellar emission is compared with {\it (i)} a model assuming strong protostellar X-ray emission and {\it (ii)} a two-dimensional (2D) model including emission in the far UV (FUV, 6 -- 13.6 eV) irradiating the outflow walls that separate the outflowing gas and infalling envelope material. We confirm that the effect of FUV in two dimensional models with enlarged irradiated surfaces is clearly noticeable. A molecule that is very sensitive to FUV irradiation is CH$^+$, enhanced in abundance by more than 5 orders of magnitude. The HIFI observations of CH$^+$ lines agree with the two-dimensional FUV model by Bruderer et al. which computes abundances, non-LTE excitation and line radiative transfer.$^{20}$ It is concluded that CH$^+$ is a good FUV tracer in star-forming regions. The effect of potential X-ray irradiation is not excluded, but cannot be demonstrated by the present data.
\end{abstract}

%%%%%%%%%%%%%%%%%%%%%%%%%%%%%%%%%%%%%%%%%%%%%%%%%%%%%%%%%%%%%%%%%%%%%
%% Start the main part of the manuscript here.
%%%%%%%%%%%%%%%%%%%%%%%%%%%%%%%%%%%%%%%%%%%%%%%%%%%%%%%%%%%%%%%%%%%%%
\section{Introduction}

Hydrogen is by far the most abundant element in the universe and plays the dominant part in the chemistry of star-forming regions. The next frequent element apart of helium is oxygen, which is more than a thousand times less abundant by number, followed by carbon, neon, nitrogen, magnesium, silicon, iron, and sulfur.$^{1}$ Mg, Si, and Fe are trapped in dust grains. Thus H, O, C, N, and S are the major chemical players.  Molecules of the form HX are known as hydrides, where X is a heavier element, an ion, or another hydride. Hydrides are the most basic molecules through which astrochemistry develops further. They are of fundamental importance for astrochemistry (reviews e.g. in Refs. 2 and 3).

The formation of hydrides depends on abundances, temperature, activation energy, irradiation, and first ionization potential of X.$^{4}$ The specifics have to be studied for each molecule, but some general comments are possible. Hydrogen abstraction
\begin{equation}
{\rm X + H_2 \rightarrow HX + H}
\label{equ1}
\end{equation}
is one of the typical reactions, but often has an activation energy of several thousand degrees (UMIST$^{5}$). Such reactions occur therefore only at high gas temperatures. More important for the chemistry in temperature regions less than a few hundred degrees is protonation through
\begin{equation}
{\rm X + H_3^+ \rightarrow HX^+ + H_2}
\label{equ2}
\end{equation}
(e.g. Ref. 6). In general, this reaction needs no activation energy and takes place also in low-temperature regions of the envelope of an accreting protostar. This makes H$_3^+$ a key species in ionization driven chemistry.$^{7,8}$ In star-forming regions internal sources of extremely intense ionizing radiation are expected. Cosmic rays, UV radiation above 6 eV, and X-rays ionize molecules and dissociate them to atoms and ions. Ionizing radiation also heats irradiated regions through the photoelectric effect on dust grains. The density in the envelope of protostars is of the order of 10$^4$ to $10^7$ cm$^{-3}$ and can exceed $10^{13}$ cm$^{-3}$ in accretion disks. At these densities ion molecule reactions can play an important role even for the formation of neutral hydrides. Reactions of the type (\ref{equ2}) dominate the formation of OH$^+$, CH$^+$, and SH$^+$ in the irradiated parts of star-forming regions that are cooler than a few 100 K. Thus hydrides, and in particular their ions, are expected to be tracers of ionizing radiation.

Hydrides are well-known in interstellar diffuse clouds, where the general interstellar FUV radiation field and cosmic rays ionize low-density gas and drive the chemistry. The first molecules in space, CH and CH$^+$, were detected already in the late 1930s in absorption by the diffuse interstellar medium. Hydride lines in absorption by diffuse clouds were extensively observed by Herschel/HIFI.$^{9-11}$ Here we concentrate on star-forming regions where the gas is denser and hotter and may be intensely irradiated by the nearby protostar. The difference can be remarkable, as for instance CH$^+$ in diffuse clouds cannot be modeled by external irradiation alone.$^{12}$ At the higher radiation density in star-forming regions, however, modelers have successfully reproduced the observations without additional energy input.$^{13-15}$

\rm In addition to chemistry, ionized molecules play an important role in the physics of star and planet formation. Ionized molecules, ions, and electrons couple the gas to the magnetic field. The magnetic field dominates the gas motions in the inner part of star-forming regions, funnels accretion, accelerates outflows and jets in the rotation axis perpendicular to the accretion disk, and causes viscosity in the disk. As long as the young stellar objects are deeply imbedded, none of the ionizing radiation escapes the envelope and can be observed;$^{16}$ the physics of the early phase of star and planet formation is still unclear. Chemistry can help to characterize these processes.

X-rays $>$ 1 keV are produced by magnetic activity or shocks. They are absorbed mostly by the gas after a half-power hydrogen column density of some $10^{23}$ cm$^{-2}$; beyond 20 keV this value is an order of magnitude larger. X-rays penetrate accreting envelopes and disks without scattering.  X-rays (and cosmic rays) can ionize all atoms and molecules, in particular O, N, and F, having a first ionization potential above 13.6 eV, the value for hydrogen atoms.

Internal extreme UV radiation (EUV, $>$13.6 eV) is expected from the surface of a massive star, or from accretion and outflow shocks. It is effectively absorbed by hydrogen ionization, causing an inner hole in the envelope around a young stellar object (compact HII region). Thus FUV, at lower energies, irradiates exposed surfaces such as the inner edge of the envelope, surfaces of the accretion disk, and the walls carved out by the outflows. From the surface, the FUV photons scatter on dust into the dense gas. The FUV irradiated border region of the infalling envelope has a hydrogen column density of a few times 2$\cdot 10^{21}$ cm$^{-2}$, the half-power FUV penetration depth $N_H^p$, where FUV is absorbed by dust.  We note that the FUV luminosity for example in W3 IRS5, a cluster of high-mass objects, is $3\cdot 10^{38}$ erg s$^{-1}$ compared to $5\cdot 10^{30}$ erg s$^{-1}$ in X-rays. The ratio is expected to be less pronounced in low-mass objects where the stellar surface does not emit FUV, but emission by accretion shocks is poorly characterized. Thus X-rays may play a minor role except in high-density regions near protostars, such as protoplanetary disks, where FUV does not penetrate.

We present results of the subprogram `Radiation Diagnostics' of the Herschel Key Program `Water in Star-forming regions with Herschel' (WISH$^{17}$). The goal of the subprogram is to explore the possibilities of identifying FUV and X-ray emission through chemistry in deeply embedded objects and to quantify such high-energy radiation and their influence on H$_2$O in star and planet-forming regions. Here we compare the efficiencies of the two ionization sources, FUV vs. X-rays for the case of two high-mass star-forming regions.

\begin{table}[htb]
\begin{center}
\resizebox{16cm}{!}{
\begin{tabular}{lcrcc|rrrrc}   % c = center, l = left justified, r=right just.
\hline \hline
Species &Tran-& Frequency&$E_u$&$A_{ul}$& \multicolumn{2}{c}{W3 IRS5}& \multicolumn{2}{c}{AFGL 2591}&Line \\
 & sition& [GHz]&[K]&[s$^{-1}$]&$N_i$ [cm$^{-2}$]&$T_{ex}$ [K]&$N_i$ [cm$^{-2}$]&$T_{ex}$ [K]&mode \\
\hline\\
OH$^+$&$1_1-0_1$&1033.1186$^a$ & 49.58& 1.8(-2)&2.1(14)&9&7.7(13)$^h$&13.2&abs\\
OH$^+$&$2_1-1_1$ &1892.2271$^a$ & 140.4& 5.9(-2) &&&&&\\
CH$^+$&1 - 0&835.1375$^{a,b}$&40.08&6.4(-3)&9.6(12)$^h$&43$^h$&8.5(12)$^h$&38$^h$&em\\
CH$^+$&2 - 1&1669.2813$^a$&120.19&6.1(-2)&4.4(13)$^h$&2.7$^h$&1.2(14)$^h$&8$^h$&abs\\
NH$^+$&$1_{\frac{3}{2}-}-1_{\frac{1}{2}+}$&1012.5400$^c$& 48.59& 5.4(-2)&$<$4.0(10)&40&$<$3.4(10)&40&em\\
NH$^+$&$1_{\frac{3}{2}+}-1_{\frac{1}{2}-}$&1019.2107$^c$& 48.91& 5.5(-2)&$<$9.5(11)&9&$<$2.8(11)&9&abs\\
SH$^+$&$1_{2\frac{5}{2}}-0_{1{\frac{3}{2}}}$&526.0479$^{a,d}$& 25.25&9.7(-4)&2.5(12)&40&$<$5.5(11)&40&em\\
SH$^+$&2$_{3\frac{7}{2}}$ - 1$_{2\frac{5}{2}}$&1082.9117$^{a,d}$& 77.2&9.1(-2)&&&&&\\
SH$^+$&3$_{4\frac{9}{2}}$ - 2$_{3\frac{7}{2}}$&1632.5179$^{a,d}$& 155.6&3.1(-2)&&&&& \\
H$_2$O$^+$&$1_{10\frac{3}{2}\frac{3}{2}}-1_{01\frac{1}{2}\frac{1}{2}}$&604.6785$^{a,e}$& 59.2& 1.3(-3)&&&&&\\
H$_2$O$^+$&$2_{11}-2_{02}$&746.5456$^{a,e}$& 125.0& 1.1(-2)&&&&&\\
H$_2$O$^+$&$1_{11\frac{3}{2}}-0_{00\frac{1}{2}}$ &1115.2041$^{a,e}$ & 53.5& 3.1(-2)& 4.8(12)&9&1.7(13)$^h$&18.5&abs\\
H$_2$O$^+$&$2_{12\frac{3}{2}\frac{3}{2}}-1_{01\frac{1}{2}\frac{1}{2}}$&1638.9348$^{a,e}$& 108.8& 7.3(-2)&&&&&\\
H$_3$O$^+$&4$_{30}$ - 3$_{31}$&1031.2937$^f$&232.2 & 5.1(-3)&2.3(13)$^i$&239$^i$&4.3(12)&239&em \\
H$_3$O$^+$&4$_{20}$ - 3$_{21}$ &1069.8266$^f$&268.8&9.9(-3)&&&&& \\
H$_3$O$^+$&6$_{21}$ - 6$_{20}$ &1454.5625$^f$&692.6&7.1(-3)&&&&& \\
H$_3$O$^+$&2$_{11}$ - 2$_{10}$ &1632.0910$^f$&143.1& 1.7(-2)&&&&&\\
OH&$\frac{3}{2}+$ - $\frac{1}{2}-$ &1837.8168$^f$& 270.1& 6.4(-2)&3.9(16)$^{h,k}$&35$^{h,k}$&--&--&em\\
CH&$\frac{3}{2}_{2,-}-\frac{1}{2}_{1,+}$ &536.7611$^{a,g}$& 25.76& 6.4(-4)&9.3(13)&40&6.6(14)$^h$&22.6&em\\
CH&$\frac{5}{2}_{3,+}-\frac{3}{2}_{2,-}$ &1661.1074$^{a,g}$& 105.48& 3.8(-2)&&&&&\\
NH&$1_{2,\frac{5}{2},\frac{7}{2}}$-$0_{1,\frac{3}{2},\frac{5}{2}}$  &974.4784$^a$& 46.77& 6.9(-3)&&&&&\\
NH&$1_{1\frac{3}{2},\frac{5}{2}}$-$0_{1\frac{3}{2},\frac{5}{2}}$  &999.9734$^a$& 47.99& 5.2(-3)&1.2(14)&6&1.5(14)&6&abs\\
SH&3$_{1}$ - 2$_{-1}$&1447.0123$^f$&640.6& 8.1(-3)&$<$ 2.7(14)&150&--&--&em\\
\hline
\end{tabular}}
\end{center}
\vskip-0.1cm
\caption{Frequency, upper level energy, and Einstein A coefficient of observed lines. The numbers in parentheses give the decimal power. Molecular and atomic data are taken from: $^a$CDMS$^{40}$; $^b$Ref. 41; $^c$Ref. 42 and the hyperfine components measured by Ref. 43; $^d$Refs. 44 and 45, Einstein-A coefficient calculated from the constants by Ref. 46 and a dipole moment of 2.4 D; $^f$JPL catalogue$^{47}$; $^g$Ref. 48. The total column density, $N_i$, is given for the high-mass objects W3 IRS5 (Ref. 19  and new data) and AFGL 2591 (Ref. 31 and new data), assuming an excitation temperature $T_{ex}$ (see text). $^h$From slab model fitting; $^i$from rotational diagram; $^k$envelope component, from Ref. 21. The line mode indicates emission or absorption, no entry means no detection.}
\label{table1}
\end{table}

\section{Observations}
We report here on observations using the {\it Heterodyne Instrument for the Far-Infrared} (HIFI$^{18}$) onboard the Herschel Space Observatory of the hydrides OH, CH, NH, SH and their ions OH$^+$, CH$^+$, NH$^+$, SH$^+$, H$_2$O$^+$, and H$_3$O$^+$ in dense star-forming regions. HIFI observes in the frequency range 480 - 1910 GHz with a spectral resolving power $\lambda/\Delta \lambda$ up to $10^7$. The selection of molecules and lines is given in Table 1. We present results on two extensively observed high-mass objects, AFGL 2591, from which no X-rays have been detected, and W3 IRS5, a source of X-rays. Some of the results have already been published by other subprojects of WISH.$^{19-21}$ Here we present an overview on all lines of both sources and confront them with models of different irradiation. A summary paper including all sources is in preparation.

\begin{figure}[htb]
\centering
\resizebox{16cm}{!}{\includegraphics[angle=270]{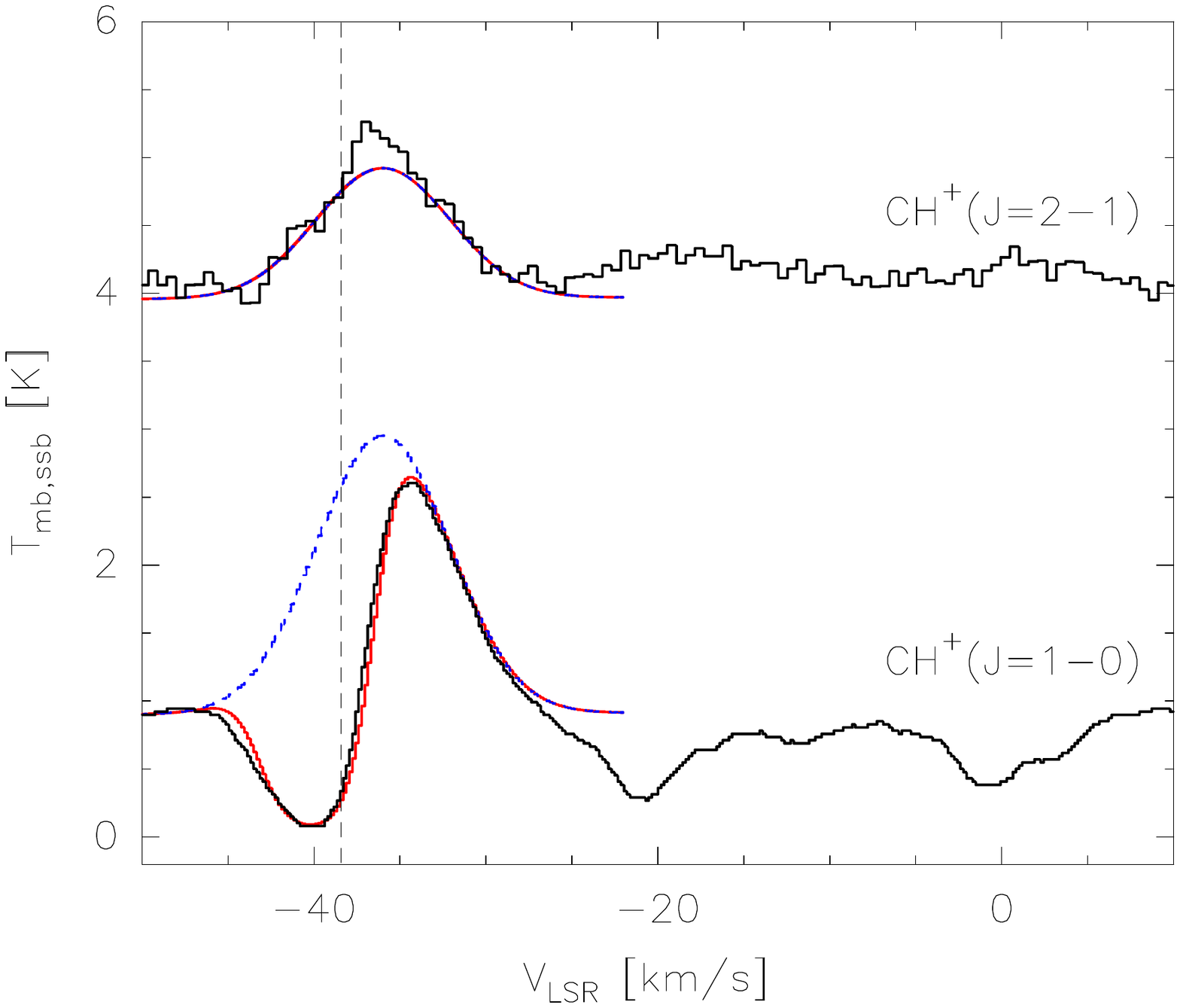}}
\caption{Line transitions of CH$^+$ observed with Herschel/HIFI towards W3 IRS5. The main beam temperature is shown; the continuum is corrected for single sideband. Frequency is presented as Doppler shift relative to the local standard of rest (LSR), where zero corresponds to the rest frequency of the respective line.  The systemic velocity of W3 IRS5, -38.4 km s$^{-1}$, is shown by a vertical dashed line. The blue dashed curve indicates the contribution of the emission component. The red curve combines the contribution of the emission and absorption components fitted by the slab model (from Ref. 22).}
\label{CH+1-0}
\end{figure}

\section{Results}

\ref{CH+1-0} presents an exemplary observation of a high-mass object. The $J=1-0$ line of CH$^+$ has a so-called P Cygni profile, produced by line emission of stationary hot gas together with cold gas moving toward us (blue shifted, negative velocity) absorbing the continuum emission of the warm dust in the central region. It is assumed that the absorption region in front of the emission region is associated with an outflow of the young stellar object (YSO). In contrast, the $J=2-1$ line shows only emission. The emission component in both lines is relatively narrow ($\Delta V \approx 7 $ km s$^{-1}$)  and symmetric, unlike emission expected from shocked regions. It is consistent with an origin in the entrained interface between the outflows and the envelope. Combining the two spectra, the column density and temperature of the emission and absorption components were estimated using a non-LTE one-dimensional slab model.$^{13}$ The best fitting models are listed in \ref{table1}. The emission peak in the $J=2-1$ transition is slightly redshifted and possibly affected by absorption on the blue side as is the case for AFGL 2591,$^{13}$ consistent with the slab modeling. Emissions and absorptions at $V_{LSR} > -25$ km s$^{-1}$ are probably caused by foreground clouds and are not modeled.

All hydrides searched for were detected, except NH$^+$ and SH (\ref{table1}). Only the line components moving within 15 km s$^{-1}$ of the systemic YSO motion are listed. The ground-state  lines of OH$^+$, H$_2$O$^+$, and NH show pure absorption, that of CH$^+$ has mixed absorptions and emission, whereas the observed transitions of OH, CH, SH$^+$, and H$_3$O$^+$ are in emission. The total column density for OH$^+$, CH$^+$, H$_2$O$^+$, CH and OH were determined from slab model fitting, for H$_3$O$^+$ from the rotational diagram. For all other molecules level column densities are extracted by integrating the observed lines or absorption profiles of a species and neglecting re-emission or re-absorption of the final state (thus optical depth $\tau \ll 1$). The level column densities were converted to total column densities summed over all levels, assuming an appropriate excitation temperature and LTE (Tab. 1). For OH$^+$ an upper limit for $T_{ex}$ of 50 K is used, derived from the non-detection of the (2-1) line. OH$^+$, NH$^+$ and H$_2$O$^+$, observed in absorption of the groundstate were assumed to have a low excitation temperature (9 K). For the column densities of NH$^+$ and SH$^+$ in emission, we assumed 40 K in analogy to the CH$^+$ slab model fits; for CH and SH also 40 K in analogy to OH. We note here that Ref. 23 assume 100 K for CH in emission for another high-mass object. This temperature would increase the CH column density by a factor of 1.6. For H$_3$O$^+$ towards W3 IRS5, the rotational diagram yields $T_{ex} = 239$ K.$^{19}$ The same excitation temperature is applied also to H$_3$O$^+$ in AFGL 2591. NH in absorption towards a low-mass source was found also by Ref. 24, reporting an excitation temperature of only 5.5 K. For the upper limit of SH in emission, a temperature of 150 K was assumed. Increasing $T_{ex}$ to 300 K would reduce the upper limit by a factor of 4.  This demonstrates the uncertainty of the derived column densities from a single line in emission, often dominated by the uncertain excitation temperature. Note that generally in cases of non-detection, the assumption of emission causes a lower upper limit.

\begin{figure}[htb]
\centering
\resizebox{8cm}{!}{\includegraphics{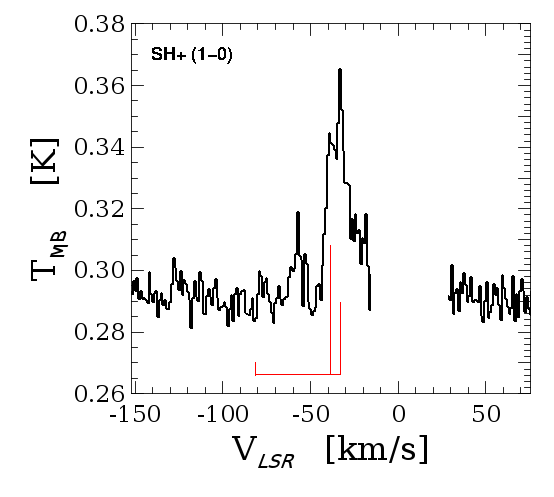}}
\caption{Line transitions of SH$^+$  discovered with Herschel/HIFI for the first time. Frequency is presented as Doppler shifted relative to the local standard of rest (LSR), where zero corresponds to 526.047947 GHz, the rest frequency of the main line. The 3 fine structure lines are indicated by vertical lines proportional in length to their theoretical intensity at 300 K. They are shifted by the systemic velocity of the observed object, W3 IRS5, amounting to -38.4 km s$^{-1}$ (adopted from Ref. 19).}
\label{SH+1-0}
\end{figure}

SH$^+$ (\ref{SH+1-0}) and H$_2$O$^+$ were detected for the first time in star-forming regions thanks to the new frequency range that has become observable with Herschel.$^{13,19,25-27}$ The new molecules were only detected in their transitions to or from the ground state, SH$^+$ in emission and H$_2$O$^+$ in absorption. SH$^+$ was detected only towards W3 IRS5, well known for its active sulfur chemistry,$^{28}$ but not in AFGL 2591.

The non-detection of SH was predicted.$^{20}$ It was not observed in a groundstate transition, as is the case for the other non-detections in Tab. 1. SH was recently found$^{29}$ in an absorbed groundstate transition towards the high-mass source W49N with a column density of $4.6\cdot 10^{12}$cm$^{-2}$.

\section{Models and Discussion}

Molecular abundances were predicted before observations, using a time-dependent chemical model that considers FUV and X-ray irradiation. Starting from a density model of AFGL 2591,$^{30}$ the local FUV intensity and dust temperature are calculated using a Monte Carlo radiative transfer code. Prestellar dense cloud molecular and atomic constituents are the initial values. The chemical evolution is then simulated$^{31}$ using the UMIST Database for Astrochemistry (updated to 2006$^{5}$) following the chemical network simulations by Refs. 32 and 33. The calculation time is significantly reduced by using the grid approach.$^{31}$ This allows to self-consistently solve for the chemistry together with the gas temperature, which is obtained from the equilibrium between heating and cooling rates similar to classical FUV irradiated models in dense interstellar clouds.$^{34}$ We assume a chemical age of 5$\cdot 10^4$ yr typical of these sources$^{8}$ and determine the abundances at all positions in the region.

We compare observations and models in two steps. In a rough first step, we estimate the column density of molecules from chemical models assuming constant beam size and a simplified geometry, and confront them with the column densities derived from observations as given in \ref{table1}. In a second step, possible only for the CH$^+$ molecule, we compare at the level of line intensities in two transitions, including full radiative transfer, frequency-dependent beam size, and geometry. The latter comes in if the emission region of most molecules considered here is a relatively thin layer (some 100 AU) on the outflow wall. Thus the column density depends on the outflow inclination and opening angle, which are not well constrained.

\subsection{Comparison of Column Densities}

The chemical model abundance $\chi_i$ of species $i$ is available from model calculations as the average over a volume of given radius $R$. In the two dimensional models the average includes the inhomogeneity due to the outflow, the inflow, and the transition zone, known as 'outflow wall'. The volume averaged abundance is generally defined as
\begin{equation}
 <\chi_i> =  {\int_{r<R} {n_i dV}\over \int_{r<R}{n_H dV}}\ \ ,
\end{equation}
where $n_i$ is the density of species $i$ ; $n_H$ is the total hydrogen density [= $n(H) + 2n(H_2)$]. The modeled molecular column densities can be approximated for a beam of radius $R$ by
\begin{equation}
 N_i =  {<\chi_i> \int_{r<R}{n_H dV}\over {\pi R^2}}\ \ .
\end{equation}
The results are presented in \ref{table2} for $R=20000$ AU, implying $\int_{r<R}n_H dV = 18.7 M_sun m_H^{-1} = 2.2\cdot 10^{58}$.

The model parameters have been studied in a range of protostellar luminosities, distances, cavity shapes and cavity densities in Ref. 20, Tab. 9. The distance to AFGL 2591 was recently measured by VLBI to 3.3 pc.$^{35}$ The radius of 20000 AU assumed here then corresponds to a beam of 12$''$, appropriate to the highest frequencies observed. A larger beam generally reduces the modeled line intensities because of beam dilution for centered emissions. The effect was shown to be less than an order of magnitude$^{20}$ and will be discussed together with other model parameters, like the unknown outflow opening angle or existence of a disk, that have effects of similar magnitude.

\begin{figure}[htb]
\centering
\resizebox{16cm}{!}{\includegraphics{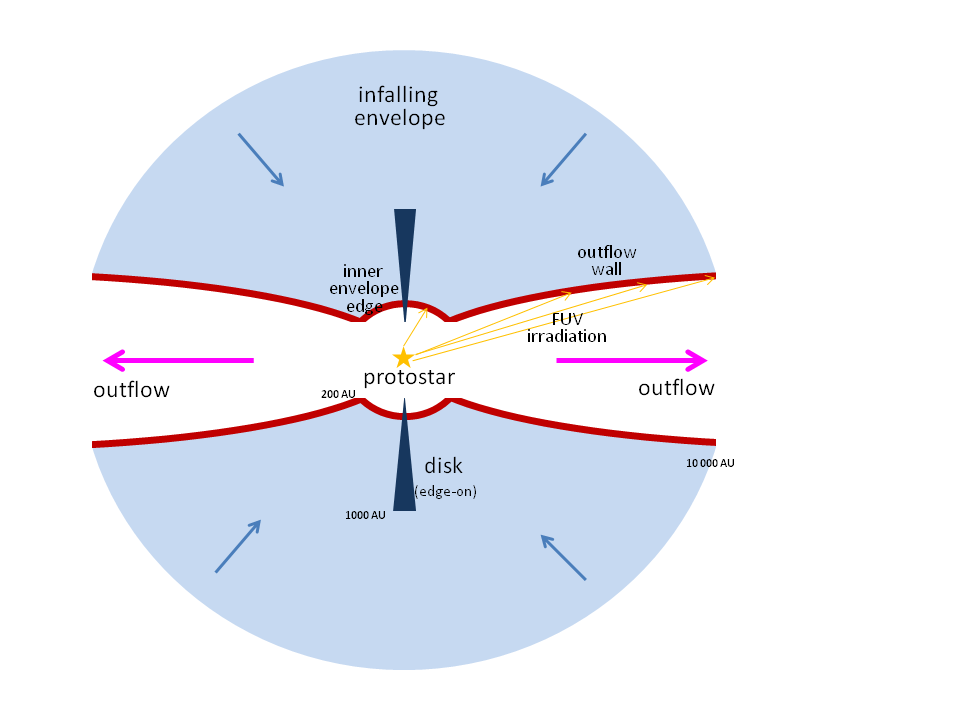}}
\caption{Scenario of star-forming region illustrating irradiation by internal FUV. The model is rotationally symmetric relative to the outflow axis indicated by two arrows. Some approximate distances are indicated in Astronomical Units (AU).}
\label{cartoon}
\end{figure}

\ref{table2} compares three models:

1. The 'noFnoX' model is a one-dimensional spherically symmetric model without high-energy irradiation except external cosmic rays. A radial dependence on density and the temperature model is used.$^{33}$ The model assumes a cavity of 200 AU radius in the center. We use the model of Ref. 20 without irradiation, which is similar within the uncertainties to model 0 by Ref. 8.

2. Model 'X' is also one-dimensional and assumes a high X-ray luminosity, $10^{32}$ erg s$^{-1}$, but no protostellar FUV emission.  There is little difference in the chemical effects between X-ray and cosmic ray irradiation.$^{31}$ The effect of protostellar X-ray irradiation decreases geometrically ($\propto r^{-2}$) and equals cosmic ray ionization in the envelope at a radius of more than 6000 AU for the above X-ray luminosity. Protostellar X-ray emission would therefore enhance ionization preferentially near the protostar.

3. St{\"a}uber et al.$^{33}$ have shown in 2007 that protostellar FUV irradiation of the inner edge of the envelope and X-ray emission in the range of observed luminosities are insufficient to explain the observed molecular abundance of CO$^+$. In a spherically symmetric FUV model CO$^+$ is enhanced in abundance by two orders of magnitude order of magnitude, but not sufficient to explain the observations. The abundances of such molecules, including CH$^+$ and OH$^+$, increase another three orders of magnitude or more if the FUV irradiated surface is enlarged. Thus Bruderer et al.$^{20}$ presented a two-dimensional model in 2010 where the FUV emission by the central object irradiates also the walls separating the two outflows and the infalling envelope with a luminosity of $7.7\cdot 10^{37}$ erg s$^{-1}$. This irradiation corresponds to a protostellar surface of a B star having a temperature of $3\cdot  10^4$ K. The gas inside the outflow has negligible column density, and the shape of the cavity allows direct irradiation of the walls. This irradiation both ionizes and heats a thin layer with a thickness of a few times the FUV penetration depth. The surface extends out to more than 30,000 AU, a radius comparable or larger than the Herschel beam. The scenario of model FUV is illustrated in \ref{cartoon}.

\begin{table}[htb]
\begin{center}
\resizebox{14cm}{!}{
\begin{tabular}{lrrrrr}   % c = center, l = left justified, r=right just.
\hline \hline
Mole- &Model& Model & Model &Ratio &Ratio \\
cule &noFnoX  & X &FUV &X/noFnoX& FUV/noFnoX \\
&[$10^{10}$ cm$^{-2}$]&[$10^{10}$ cm$^{-2}$]&[$10^{10}$ cm$^{-2}$]& & \\
\hline\\
OH$^+$&0.12&0.15&920&1.3&8000 \\
CH$^+$&0.0023&0.0028&3300&1.2&1.4$\cdot  10^6$ \\
NH$^+$&0.0075&0.010&7.7&1.4&1000  \\
SH$^+$&23&69&72&3.0&3.1  \\
H$_2$O$^+$&0.25&0.37&360&1.5&1400 \\
H$_3$O$^+$&6100&7500&2900&1.2&0.48 \\
OH&51000&70000&120000&1.4&2.4 \\
CH&320&420&4400&1.3&14 \\
NH&2600&3900&7400&1.5&1.9\\
SH&2400&1500&240&0.63&0.099 \\
\hline
\end{tabular}}
\end{center}
\caption{Volume-averaged column densities, $<N_i> $, derived from chemical model calculations for a high-mass object (AFGL 2591) in a beam with radius 20000 AU (in units of $10^{10}$ cm$^{-2}$).}
\label{table2}
\end{table}

Models X and FUV are compared to noFnoX in \ref{table2}. X-rays enhance all abundances except SH. Except for SH and H$_3$O$^+$, all hydride abundances are enhanced also by FUV, particularly the ionized species. The much more powerful FUV irradiation makes the enhancements in model FUV generally larger than in model X.

\begin{figure}[htb]
\centering
\resizebox{17cm}{!}{\includegraphics{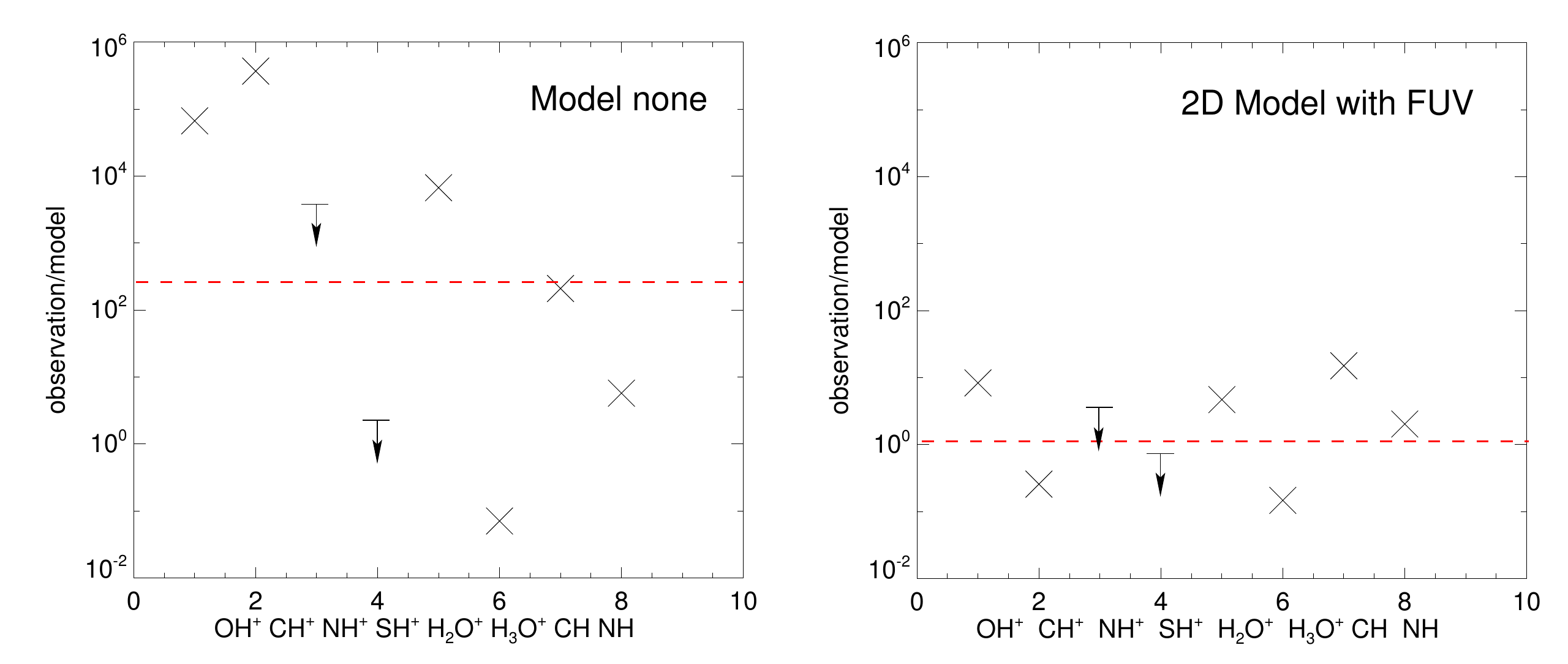}}
\caption{Comparison between observed column densities towards AFGL 2591 and theoretical values derived by Ref. 20 (\ref{table2}) from chemical models, using Eq. (4) with $a=1$. {\it Left:} model noFnoX without internal high-energy radiations; {\it right:} model FUV with internal FUV irradiation in 2D having a luminosity of 7.7$\cdot  10^{37}$ erg s$^{-1}$. The log-average value is indicated by a horizontal dashed line.}
\label{comparison}
\end{figure}

\ref{comparison} left compares observations with model predictions of beam averaged column densities in the noFnoX model. The comparison of the observations with the FUV model including 2D FUV irradiation of the outflow walls is shown in \ref{comparison} right. The two models can be compared quantitatively in the average observation-to-model ratio and its scatter over the different molecules. (i) The observed column density of CH$^+$, OH$^+$, and H$_2$O$^+$ far exceed the values given by the noFnoX model. The log-averaged observation-to-model ratio $<$log$_{10}a>$ without FUV and X-rays is 260 (linearized), neglecting upper limits. The X-ray model (not shown) reduces slightly this ratio to 220; for the FUV model it amounts to 1.06. If X-ray emission of model X is added to the FUV model, the average becomes 0.99, demonstrating that X-rays have negligible effect at the size of the {\it Herschel} beam. (ii) Furthermore, the scatter is significantly reduced by including FUV irradiation (see \ref{comparison}). In logarithmic scale, the standard deviation amounts to 2.5 for the noFnoX model, 2.4 for the X model, but diminishes to 0.65 for the FUX model. This is emphasizes the importance of internal FUV irradiation, enhancing most of the species discussed here in the outflow wall.

NH$^+$ deserves a special remark. Its upper limit for emission is far below the FUV model prediction, but not the upper limit for an absorption line at much lower excitation temperature, shown in \ref{comparison}. Reaction (\ref{equ2}) is strongly endoergic  for the production of NH$^+$ and is negligible. As N cannot be ionized by FUV, reaction (\ref{equ1}) is only competitive through X-ray or cosmic ray ionization. Therefore, high NH$^+$ abundance would suggest such irradiations. This is not the case, and in their absence NH$^+$ forms from NH by charge exchange, NH + H$^+$. There is another channel as NH has a first ionization potential of 13.49 eV, thus low enough to be ionized by FUV. However, as NH requires a high activation energy forming through reaction (\ref{equ1}), NH is very temperature sensitive and thus the resulting NH$^+$  difficult to predict.

The agreement with the predictions of the FUV model is within an order of magnitude (\ref{comparison}, right). The largest deviations are the high abundance observed for CH and low abundance for H$_3$O$^+$. We note that CH was observed at a relatively low frequency, thus with a large beam. Therefore, the deviation cannot be an effect of the assumption of a constant beam, but indicates the accuracy of this first step, limited by simplifying assumptions.

\subsection{Full Radiative Transfer Model}

In a second approach, the comparison is made at the level of line intensities, requiring radiative transfer calculations. The radiation modeling includes a non-LTE determination of the level populations, appropriate beam sizes, and the most likely outflow geometry. The excitation of the molecule by collisions as well as by the radiation field of the dust continuum and the line radiation is calculated, the radiative transfer evaluated and convolved with the Herschel beam. A severe limitation to this approach is the lack of data on the collisional cross-sections for most of the observed molecules. Bruderer et al. calculated in 2010 the full radiative model for the CH$^+$ line emission of AFGL 2521 in the Herschel beam at 835 GHz and 1669 GHz (26.3'' and 12.7'' FWHM, respectively).$^{20}$ The model assumes FUV irradiation in a two-dimensional geometry. The result for the line intensities of CH$^+$ $J=1-0$ and $J=2-1$ is close to the observed values (\ref{table_comparison}). The model was calculated for distances of 1 and 2 kpc. Increasing the distance by a factor of 2 reduced the line intensities by a factor of 2.7, resp. 2.9. The larger distance recently observed$^{35}$  reduces further the modeled values given in \ref{table_comparison} and improves the agreement.

The approach does not require the assumption of excitation temperatures and is considerably more realistic than comparing column densities. Nevertheless, there remain uncertainties concerning ({\it i}) the chemical reaction rates,$^{32}$ ({\it ii})  the difficulties in computing gas temperatures in FUV irradiated regions,$^{36}$ ({\it iii}) the absorption by cooler gas in the outer envelope or in front of the object (Fig. 1), and ({\it iv}) the geometrical and other model assumptions. Considering these unknowns, even detailed modeling remains uncertain to within a factor of a few; the agreement is well within this range. The effect of protostellar X-rays is small in view of these uncertainties.

\begin{table}[htb]
\begin{center}
\resizebox{12cm}{!}{
\begin{tabular}{lrrr}   % c = center, l = left justified, r=right just.
\hline \hline
&Abundance&Line intensity (J=1-0)&Line intensity (J=2-1)  \\
&of CH$^+$&at 835.1375 GHz&at 1669.2813 GHz\\
&relative to H&[K km s$^{-1}$] &[K km s$^{-1}$]\\
\hline\\
Model noFnoX&$5\cdot  10^{-16}$&$\ll 0.001$ &$\ll 0.001$\\
Model FUV& $4\cdot  10^{-10}$&4.3&8.9\\
Observed& $0.9\cdot  10^{-10}$&0.91$\pm$0.03&3.7$\pm$0.2 \\
\hline
\end{tabular}}
\end{center}
\caption{Comparison of AFGL 2591 chemical and radiation transfer simulations$^{20}$ assuming standard parameters and a distance of 2 kpc with Herschel/HIFI observations.$^{13}$}
\label{table_comparison}
\end{table}

\section{Conclusions}

Hydrides and in particular their ionized versions have previously been considered as probes of the low-density cool diffuse interstellar medium, as collisions with hydrogen molecules generally reduce their abundance. Indeed, spherically symmetric models of star-forming regions with internal irradiation sources predicted low abundances in the past. Such models successfully reproduced the abundance of neutral molecules in the outer envelope of AFGL 2591.$^{32}$ However, observations  by HIFI onboard {\it Herschel} presented here reveal that hydrides, and particularly the ionized ones, are more abundant than previously expected in some inner parts of star-forming regions. Here we have tested the hypothesis that the high abundance of hydrides originate from protostellar irradiation by FUV or X-rays. We compared the observations with three models, one without irradiation, one assuming a high protostellar X-ray irradiation, and one with FUV irradiation as expected from the hot protostellar surface or accretion shocks.

X-ray irradiation is not scattered, but penetrates straight into regions of higher density such as the protoplanetary disk. Its effect is limited by the generally much lower luminosity compared to FUV and by geometrical dilution. X-rays cannot be excluded. However, models assuming only X-rays underestimate on the average significantly the abundance of hydrides in high-mass star-forming regions observed here and scatter in the observations-to-model ratio more than the model that assumes FUV irradiation.

The effect of FUV irradiation is strongly enhanced by a two-dimensional geometry that increases the irradiated surface area.$^{20}$ FUV scattering produces a layer of enhanced hydride abundance with a thickness of a few FUV penetration lengths. The thickness does not depend much on the inclination angle of the irradiation at the surface. In these regions, the huge impact of FUV on the hydride chemistry in high-mass star forming regions is not only caused by ionization, but also the result of substantial gas heating by the photoelectric effect on grains of FUV radiation. The 2D geometry is essential to reach a good fit with observations.

An order-of-magnitude agreement is already apparent in the simplified approach neglecting the details of line excitation, beam size, and propagation (\ref{comparison}). In the full model including the chemical network, excitation and radiation transfer, an agreement within the smaller range of uncertainties is possible without assuming additional X-ray emission (\ref{table_comparison}).

FUV irradiation can be readily identified by molecular tracers. In star forming regions, CH$^+$ and OH$^+$ are very FUV-sensitive (\ref{comparison}). CH$^+$ emission is more enhanced and probes primarily the irradiated walls of the outflows that carve out the envelope of star-forming regions. If observable, H$_3^+$ remains the best direct tracer of high energy photons and cosmic rays. Yet, there is no clear X-ray tracer among the species considered here.

Ionized hydrides are chemically active and can drive substantial chemical evolution. If their chemistry and excitation is understood, they may become valuable probes of warm and ionized gas in the embedded phase in star and planet formation. OH$^+$, $^{13}$CH$^+$, and SH$^+$ can be observed from the ground,$^{38,39}$ and this will become particularly interesting for high spatial resolution observations with the Atacama Large Millimeter/submillimeter Array (ALMA) telescope. Given more accurate molecular collisional cross-sections and better geometrical constraints, more accurate modeling will be possible. Spatial resolution is important concerning the hot inner part of the star-forming region where the hydride lines in emission originate, as well as for the cooler regions farther out producing the enigmatic absorptions. This will allow to pose more quantitative questions about the strength of the irradiation sources.

\subsection{Acknowledgements}
We thank Takeshi Oka for his ground-braking work on the spectroscopy and astrochemistry of H$_3^+$ and other small hydrides. Martin Melchior helped with data analysis software. Floris F. S. van der Tak and an unknown referee contributed helpful criticism. We acknowledge inspiring discussions and support by the WISH team. HIFI has been designed and built by a consortium of institutes and university departments from across Europe, Canada and the United States under the leadership of SRON Netherlands Institute for Space Research Groningen. This program is made possible thanks to time guaranteed by a hardware contribution funded by Swiss PRODEX (grant 13911/99/NL/SFe). The work on star formation at ETH Zurich was partially supported by the Swiss National Science Foundation (grants 20-113556 and 200020-121676).

%%-----------------------------
%%      your bibliography
%%-----------------------------

\vfill
\eject
\begin{figure}[htb]
\centering
\resizebox{5.7cm}{!}{\includegraphics{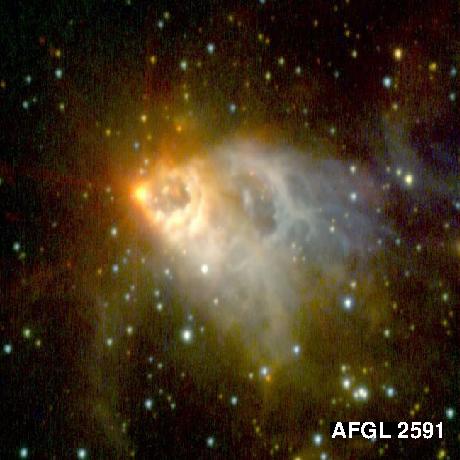}}
\caption{For Table of Contents}
\label{cartoon}
\end{figure}

\end{document}